\documentclass[conference]{IEEEtran}

\ifCLASSINFOpdf
  \usepackage[pdftex]{graphicx}
\else
  \usepackage[dvips]{graphicx}
\fi
\usepackage{subcaption}
\usepackage{textcomp}
\usepackage{amsmath}

\usepackage{stfloats}
\usepackage{algorithm,algorithmic}
\usepackage{multirow}
\usepackage{booktabs}

\makeatletter 
\newcommand{\linebreakand}{%
  \end{@IEEEauthorhalign}
  \hfill\mbox{}\par
  \mbox{}\hfill\begin{@IEEEauthorhalign}
}
\makeatother 

\begin{document}

\title{Insonification Angle-based Ultrasound Volume Reconstruction for Spine Intervention}

\author{
\IEEEauthorblockN{Baichuan Jiang}
\IEEEauthorblockA{Department of Computer Science\\
Johns Hopkins University\\
Baltimore, Maryland 21218\\
Email: baichuan@jhu.edu}
\and
\IEEEauthorblockN{Keshuai Xu}
\IEEEauthorblockA{Department of Computer Science\\
Johns Hopkins University\\
Baltimore, Maryland 21218\\
Email: keshuai@jhu.edu}
\and
\IEEEauthorblockN{Abhay Moghekar}
\IEEEauthorblockA{Department of Neurology\\
Johns Hopkins Medical Institute\\
Baltimore, Maryland 21205\\
Email: am@jhmi.edu}
\and

\linebreakand

\IEEEauthorblockN{Peter Kazanzides}
\IEEEauthorblockA{Department of Computer Science\\
Johns Hopkins University\\
Baltimore, Maryland 21218\\
Email: pkaz@jhu.edu}
\and
\IEEEauthorblockN{Emad M. Boctor}
\IEEEauthorblockA{Department of Computer Science\\
Johns Hopkins University\\
Baltimore, Maryland 21218\\
Email: eboctor@jhu.edu}
}

\maketitle

\begin{abstract}
Ultrasound-guided spine interventions, such as lumbar-puncture procedures, often suffer from the reduced visibility of key anatomical features such as the inter-spinous process space, due to the complex shape of the self-shadowing vertebra. Therefore, we propose to design a wearable 3D ultrasound device capable of imaging the vertebra from multiple insonification angles to improve the 3D bone surface visualization for interventional guidance. In this work, we aim to equip the imaging platform with a reconstruction algorithm taking advantage of the redundant ultrasound beam angles. Specifically, we try to weight each beam’s contribution for the same reconstructed voxel during the reconstruction process based on its incidence angle to the estimated bone surface. To validate our approach, we acquired multi-angle ultrasound image data on a spine phantom with a tracked phased array transducer. The results show that with the proposed method the bone surface contrast can be significantly enhanced, providing clearer visual guidance for the clinician to perform spine intervention.

\end{abstract}


\IEEEpeerreviewmaketitle

\section{Introduction}
Lumbar puncture is one of the most commonly performed spinal interventions with over 360,000 procedures performed every year in emergency departments alone within the United States and still increasing \cite{Vick2018Epi,Kroll2015Trends}. It can be particularly challenging for obese or pregnant patients who can hardly bend over to open up the inter-spinous gap, which likely results in missing the initial needle placement and causing iatrogenic complications such as hematoma and nerve damage \cite{Brown2016Iatro}. Therefore, extensive research has been conducted on utilizing intra-operative ultrasound imaging for lumbar puncture \cite{Soni2016} because it is able to provide radiation-free and real-time interventional guidance. However, it often suffers from the poor visualization of key anatomical structures such as the bone boundaries of the inter-spinous space, because the complex shapes of the self-shadowing lumbar vertebrae cannot be easily visualized by ultrasound imaging from a single insonification angle. Therefore, we have proposed a multi-angle imaging setup ``AutoInFocus" using a wearable ``patch-like" ultrasound device that aims to improve the key structure visibility for US-guided spine intervention \cite{Xu2022Auto}. 

\begin{figure*}[!t]
\centering
\includegraphics[width=7in]{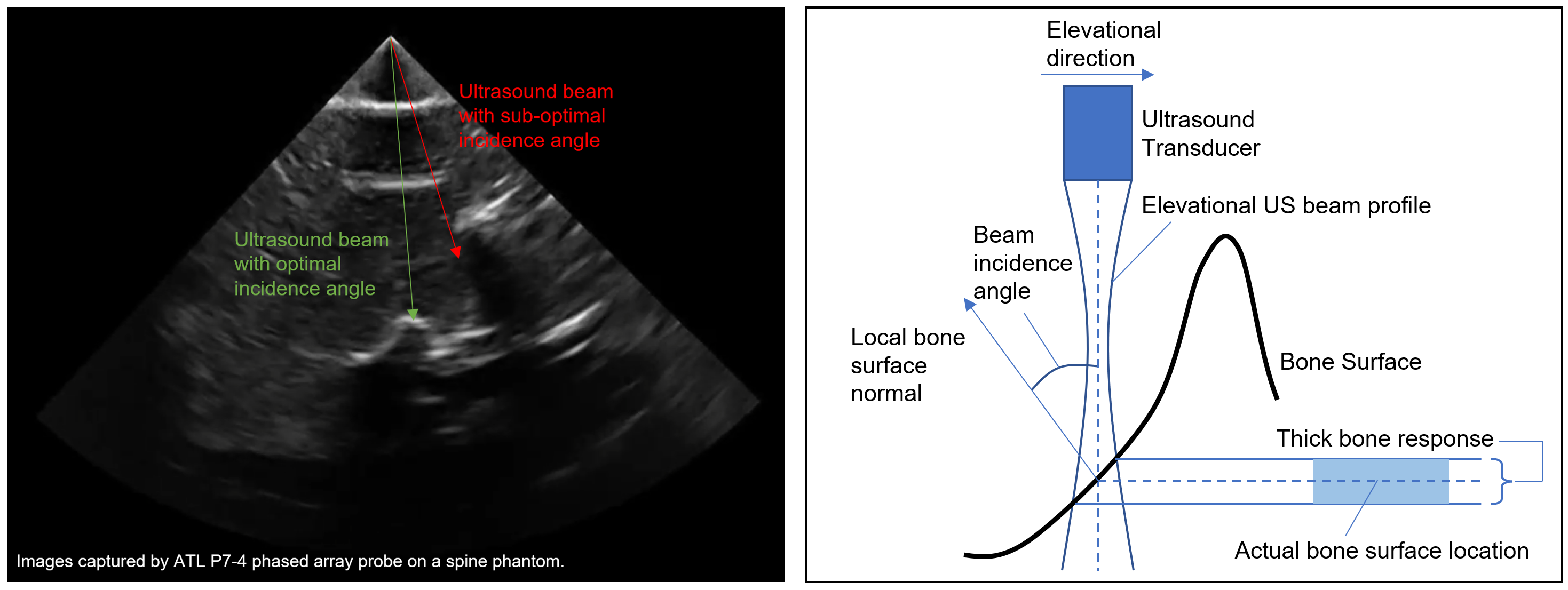}
\caption{The effects of insonification angle for spinal ultrasound imaging. \textit{Left}: comparison between different beam insonification angles on bone surfaces using an actual ultrasound image. \textit{Right}: schematic diagram demonstrating the early-and-late echo artifacts and their relation to the beam angle of incidence.}
\label{fig_anlge_effect}
\end{figure*}

There are three main advantages of having diverse insonification angles when imaging complex-shaped bone anatomy such as the lumbar vertebrae as depicted in Figure \ref{fig_anlge_effect}. First, the bone has high acoustic attenuation, which results in significant shadowing artefacts that block out important anatomical features within the shadow region. Having diverse insonification angles can reduce the blind area and reveal the true anatomical structures for the maximum extent. Second, the reflection energy from the bone surface is highly correlated with the ultrasound beam incidence angle.  As shown in this image, surfaces hit by ultrasound beams perpendicularly will have much stronger reflections to be captured by the ultrasound transducer, whereas US beams with large angles of incidence will present minimal visualization for the surfaces. Therefore, with multiple imaging angles, the chance of bone surfaces being hit by optimal beam angles will be increased. Second, because of the elevational beam thickness, early-and-late-echo imaging artefacts can be observed if the beam is not perpendicular to the bone surface, giving the bone surface a thick and saturated appearance in the image. With optimal beam angles, the surface response can be sharper and clearer for better interventional guidance accuracy. 

Various methods have been proposed to reconstruct a 3D volume from a set of tracked 2D ultrasound slices, including the voxel-based method, pixel-based method and function-based method \cite{Solberg2007Freehand}. However, currently available methods do not incorporate patient-specific anatomical information during the reconstruction process. In this work, we will present an algorithm that can incorporate the estimated bone surface information and take advantage of the diverse insonification angles from the imaging data to reconstruct an ultrasound spine volume with superior bone surface clarity. 

\section{Methodology}
The key for the algorithm is taking advantage of the redundant ultrasound beams available. “Redundant” beams refers to the fact that within the workspace of a motorized phased array probe, the same point on the spine surface can possibly be hit by ultrasound beams multiple times from different kinematics configurations. Therefore, we try to weight each beam’s contribution to the same reconstructed voxel during the reconstruction process based on its incidence angle to the estimated bone surface. Next, we will introduce the algorithm input preparation followed by the reconstruction process. 

\subsection{Reconstruction algorithm input preparation}
Three inputs are required for running our proposed reconstruction algorithm:
\subsubsection{Tracked US images}
To reconstruct a 3D volume from a set of 2D ultrasound images, the spatial transformation of each image slice will be required. The image pixel spacing in millimeters also needs to be known and applied to the images for correctly locating each pixel in 3D space. 
\subsubsection{Bone surface probability map}
To compute the angle-based weighting factor, we will need a scalar volume map that represents the probability of a voxel located on the bone surface. In practice, this volume map can be the output of a deep learning-based bone surface estimator. In this work, to evaluate the effectiveness of the reconstruction algorithm by itself, we assume we already have a decent surface estimator that we emulated by post-processing a registered CT model to represent estimator output. We first converted the lumbar CT segmentation into a 3D label map, and then used a 3D Sobel filter to extract bone boundaries. Second, we applied a 3D Gaussian filter with 10 voxel standard deviation to simulate the uncertainty in estimation, and in the end we max-normalized the intensities between 0 and 1 to represent the bone probability. 
\subsubsection{Bone surface gradient map}
With the 3D surface probability map, we can estimate the surface normal orientation by computing the 3D gradient of the surface probability. We first apply a Sobel filter in the X, Y and Z directions separately to obtain gradient values for each direction at each voxel. Then we stack the scalar values as a 3D gradient vector for each voxel and normalize. In this way, we can compute a map that contains a normalized 3-dimensional vector that represents the reverse of the bone surface normal direction. 

\subsection{Insonification angle-based volume reconstruction}
The backbone of the volume reconstruction algorithm in this work is a Pixel-based method (PBM) as introduced in \cite{Solberg2007Freehand}, which contains a distribution step (DS) followed by a hole-filling step (HFS). In the DS, we try to align the ultrasound image to the global reference coordinates based on its tracked transformation. Specifically, we apply the pixel nearest neighbor as the distribution method, such that the pixel intensity is assigned to the nearest voxel in the reconstruction volume. In the HFS, empty voxels will also be filled with nearest neighbor values up to a limit distance, such that if the gap is larger than the threshold limit we determine that the space is never visited by the ultrasound scan. 

The key difference of the algorithms introduced and compared in this work is the way to handle multiple contributions to the same voxel in the distribution step. For our baseline method, we simply average the contributions to the voxel as the final result of the DS stage, whereas in the our ``AIF-recon" algorithm, the contributions from different beams are weighted by their incidence angle quality. To better introduce our proposed algorithm, we have listed a few important variables and their explanations below: 

\begin{itemize}
  \item $P$: Tracked ultrasound image set with $N$ images of width $W$ and height $H$.
  \item $T$: Recorded transformation matrices for the tracked ultrasound images.
  \item $V_{recon}$: Target reconstruction volume of size $Sx \times Sy \times Sz$ 
  \item $V_{prob}$: Estimated bone probability map (a scalar volume) of size $Sx \times Sy \times Sz$ 
  \item $V_{dir}$: Bone surface gradient map (a vector volume) of size $Sx \times Sy \times Sz \times 3$ 
  \item $V_{count}$: Beam visiting score volume (a scalar volume) counting how much good insonification have been received at each voxel. Volume size $Sx \times Sy \times Sz$.
  \item $p_{dir}$: The ultrasound beam direction map which can be pre-computed when using a fixed imaging depth. We can compute for each pixel the direction of the beam going through the pixel. Map size $W \times H \times 3$ (Z-dimension is padded with 1's). 
  \item $p_{data}$: The scalar intensity value of a pixel to fill in the nearest voxel. 
  \item $v_{temp}$: The temporary value to be assigned to the reconstructed voxel. 
  \item $\alpha $: The empirical enhancement factor for controlling the level of enhancement (saturation) for the bone response (chosen as 0.1). 
  \item $\beta $: The empirical incidence angle cosine value threshold for applying energy compensation. 
  \item $bool_{ext}$: the boolean variable controlling whether energy compensation is active.
\end{itemize}

Given the above variable definitions, the reconstruction algorithm is shown in Algorithm \ref{algo_aifrecon} (the loop index is flattened for clarity). A few key points: 
\begin{itemize}
  \item In step 4, we first find out the correspondence between the image pixel index and the nearest reconstructed volume voxel index for distributing the intensity value.
  \item In step 7, we compute the angle weight $W_{angle}$, which is the dot product between the transformed US scanline beam direction and the surface gradient. Intuitively, if the scanline is perpendicular to the surface, is should be in line with the surface gradient and the dot product will be 1. Otherwise it is less than 1. 
  \item In step 14, we assign the bone weight $W_{bone}$ as the probability of the voxel located on the bone surface. 
  \item In step 15, we compute a temporary voxel value by multiplying the original pixel intensity with both angle weight and bone weight as well as an enhancement factor $\alpha$, and then we add this value to the original pixel intensity. This will basically enhance the bone surfaces with high bone weight and high angle weight, while keeping it the same for other voxels not on bone surfaces. 
  \item In step 19, we use a running average scheme so that we can update the current voxel value based on how often this voxel has already been visited by good input frames.
  \item In step 20, we update the visiting count not simply adding 1 each time, but actually using bone weight and angle weight so that frames with higher quality will have much larger impact on the reconstructed voxel intensity. 
  
\end{itemize}

 \begin{algorithm}
 \caption{AIF-recon algorithm}
 \begin{algorithmic}[1]
 \renewcommand{\algorithmicrequire}{\textbf{Input:}}
 \renewcommand{\algorithmicensure}{\textbf{Output:}}
 \REQUIRE $P, T, V_{prob}, V_{dir}, bool_{ext}$
 \ENSURE  $V_{recon}$
 \\ \textit{Initialisation}: $V_{recon}=\textbf{0}, V_{count}=\textbf{0}$
  \STATE Distribution Step (DS): assign image intensities to reconstruction volume
 \\ \textit{Loop through each image and each pixel:}
  \FOR {$i = 1, 2, ..., N$}
      \FOR {$j = 1, 2, ..., H\times W$}
      \STATE $idx =$ FindVoxelIndex$(i,j,T)$ 
      \STATE $p_{data} = P[i][j] $
      \STATE $p_{dirj} = p_{dir}[j]$
      \STATE $W_{angle}=p_{dirj}\cdot V_{dir}[idx]$
      \IF {($W_{angle} < 0$)} \STATE $W_{angle} = 0$ \ENDIF
      \IF {($bool_{ext}\ \text{is}\  \textit{TRUE} $)}
      \STATE $W_{angle}[W_{angle}>\beta] = 1 / W_{angle}[W_{angle}>\beta]  $
      \ENDIF
      \STATE $W_{bone}=V_{prob}[idx]$
      \STATE $v_{temp}=p_{data}\cdot W_{angle}\cdot W_{bone}\cdot \alpha + p_{data} $
      \IF {($v_{temp} > 255$)} \STATE $v_{temp} = 255$ \ENDIF
      \STATE $V_{recon}[idx]= [V_{count}[idx]/(V_{count}[idx]+1)$ 
      \\ $\cdot V_{recon}[idx]  + v_{temp} \cdot [1/(V_{count}[idx]+1)]$
      \STATE $V_{count}[idx] = V_{count}[idx] + W_{angle}\cdot W_{bone} $
      \ENDFOR
  \ENDFOR
  \STATE Hole-filling Step (HFS): fill gaps in between image frames
  \STATE $V_{recon}=$ fillGaps$(V_{recon})$ 
 \RETURN $V_{recon}$ 
 \end{algorithmic} 
 \label{algo_aifrecon}
 \end{algorithm}
 
The AIF-recon algorithm only enhances the bone surface voxels with good beam angles. To leverage the available anatomical information to the maximum extent from the surface estimation, an energy compensation framework is added so that surface voxels reached by sub-optimal beams can be compensated. Here, we utilize an acoustic reflection energy model as introduced in \cite{Burger2012Real} with equations as follows:

\begin{equation}
I_r \approx \bigg | \cos(\theta) \cdot \bigg (\frac{Z_2-Z_1}{Z_2+Z_1}\bigg )^2 \cdot I_i \bigg |
\end{equation}
\begin{equation}
I_{c} \approx \bigg | 1 \cdot \bigg (\frac{Z_2-Z_1}{Z_2+Z_1}\bigg )^2 \cdot I_i \bigg | \approx I_r \cdot \frac{1}{\cos(\theta)}
\end{equation}

Assume the incidence beam energy is $I_i$ and the reflected beam energy is $I_r$. The incidence angle is denoted as $\theta$. $Z_1$ and $Z_2$ are the acoustic impedance of the materials on each side of the interface. The ideal reflection happens when the incidence angle is 0, which is perpendicular to the surface and $\cos(\theta)$ is 1. So to recover this energy for the non-optimal beams, it means that we will need to take the reciprocal of the $\cos(\theta)$ to multiply with the original intensity to get the compensated ideal intensity $I_c$. This extension corresponds to Step 12 in Algorithm \ref{algo_aifrecon} when setting $bool_{ext}$ to $\textit{TRUE}$ in the input. Taking the reciprocal of angle weights means smaller values will actually become larger. For stability, we also included a threshold $\beta$. For $\beta = 0.1$, it means that incoming beams incident at larger than 84.3 degrees angle will not be compensated.

\section{Phantom experiment and result}
To evaluate the proposed method with multi-angle imaging data, we have set up an Atracsys\textregistered \ optical tracker to track the markers fixed on an ATL P7-4 phased array transducer for data collection. A series of calibration experiments were conducted, including speed-of-sound calibration, tracker-ultrasound temporal offset calibration, and marker-to-US spatial calibration (hand-eye calibration). Then, we used this tracked probe to scan a CT-segmented spine phantom from multiple angles. A sample image is presented in Figure \ref{fig_anlge_effect} \textit{Left}.

\begin{figure*}[!t]
\centering
\includegraphics[width=7in]{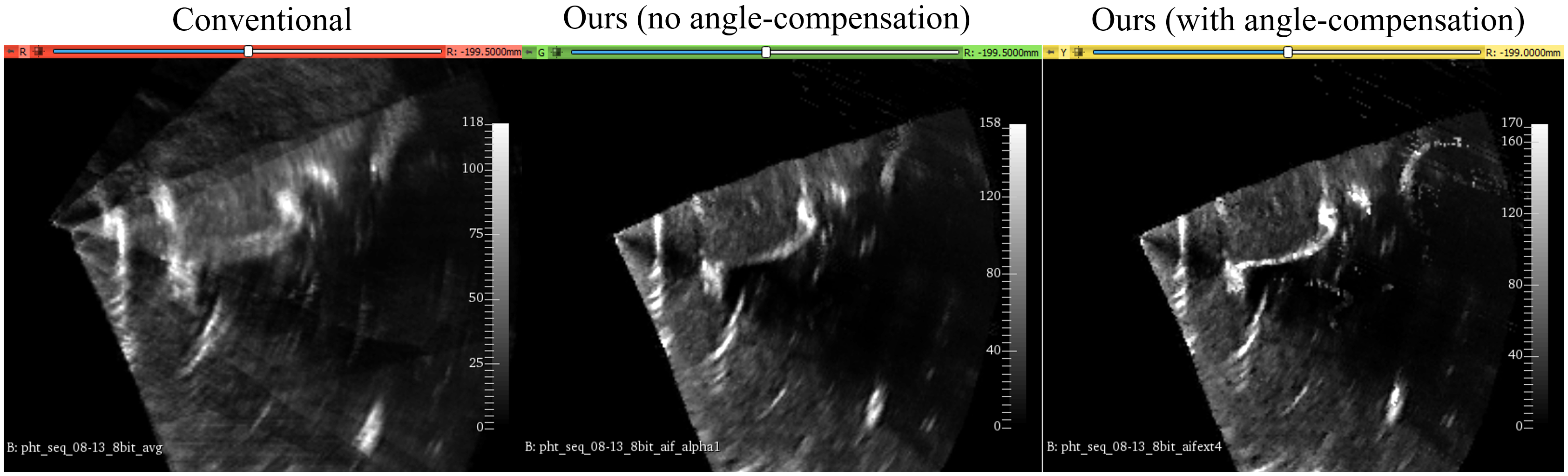}
\caption{The qualitative comparison between different reconstruction methods using cross-section view of reconstructed volumes. \textit{Left:} baseline method that averages all contributing beams for reconstructed voxel. \textit{Middle:} AIF-recon algorithm with only bone surface enhancement. \textit{Right:} AIF-recon algorithm with non-optimal beam energy compensation.}
\label{fig_qual_results}
\end{figure*}


We used the same set of images and tracking data to perform volume reconstruction via three different methods, and the results are shown in Figure \ref{fig_qual_results} (here we assume we have a decent surface estimator in the AIF-recon methods so the required inputs are simulated by post-processing the registered CT model). Our baseline method is the simple averaging of all beams visiting the same voxel, and it is shown to produce blurry boundaries. In comparison, the AIF-recon method (without angle-compensation) is a weighted average based on counted visiting scores. Therefore, for voxels where no good beams travel across, it is only taking the data from the last inserted frame, thus avoiding the blurring effect in the reconstructed data. Besides, for voxels where both good bone score and good angle score exist, it is enhancing the voxel by raising its intensity. 
In the end, we see that with our energy compensation method, the voxels that have high bone score but low angle score are compensated and now a clearer and more complete boundary is shown for the spinous process.

\section{Conclusion}
In this work, we introduced a novel volume reconstruction method with an insonification angle-based beam weighting scheme. The algorithm can take advantage of the current surface estimates and incorporate the patient-specific anatomical knowledge into the reconstruction process. Our experiment result shows that the reconstructed volume has higher contrast for the bone surface regions compared to our baseline. In addition, with an acoustic energy reflection model, we are able to compensate for the non-optimal beams and reconstruct a complete bone surface. This is the key to the success of downstream lumbar puncture guidance tasks, and we can also see the potential of this method to be applied to general spine interventions in the operating rooms \cite{Gueziri2020}.

\section*{Acknowledgment}

This study was funded by National Science Foundation SCH:CAREER Grant No. 1653322, National Institute of Health 1R43EB031731, and Analog Devices Inc. fellowship.

\end{document}